\documentclass[12pt,preprint,letterpaper,nofootinbib,superscriptaddress]{revtex4}
\pdfoutput=1
\usepackage[utf8]{inputenc}
\usepackage{epsf, array, xcolor}
\usepackage{graphicx}
\usepackage{subfigure}
\usepackage{bbold}
\usepackage{amsmath,amssymb,mathtools}
\usepackage{epigraph}
\usepackage[colorlinks=true,linktocpage=true,linkcolor=blue,citecolor=blue,urlcolor=blue]{hyperref}
\usepackage{textcomp}

\newcommand{\C}{\mathcal{C}}

\newcommand{\TphiWB}{{\C_{\phi WB}}}

\newcommand{\TW}{\C_{W}}

\newcommand{\TphiD}{\C_{\phi D}}

\newcommand{\TphiB}{\C_{\phi B}}
\newcommand{\TphiW}{\C_{\phi W}}
\newcommand{\Tphik}{\C_{\phi \square}}

\newcommand{\Tfebb}{\C_{\phi e 22}}
\newcommand{\Tfuaa}{\C_{\phi u 11 }}
\newcommand{\Tfdaa}{\C_{\phi d 11 }}
\newcommand{\TphiWBr}{{\C_{\phi WB}r}}

\usepackage{url}

\usepackage{dcolumn}% Align table columns on decimal point
\begin{document}
\pagestyle{plain}

\title{New Physics Through Drell Yan SMEFT Measurements at NLO}

\author{Sally Dawson}
\affiliation{Department of Physics, Brookhaven National Laboratory, Upton, N.Y., 11973,  U.S.A.}
\author{Pier Paolo Giardino}
\affiliation{\mbox{Instituto Galego de F\'isica de Altas Enerx\'ias, Universidade de Santiago de Compostela,}\\ \mbox{15782 Santiago de Compostela, Galicia, Spain}}

\date{\today \vspace{0.5cm}}

\begin{abstract}
Drell Yan production is a sensitive probe of new physics and as such has been calculated to high order
in both the electroweak and QCD sectors of the Standard Model, allowing for precision comparisons between theory and data.  Here we extend these calculations to the Standard Model Effective Field Theory  (SMEFT) and
present the NLO QCD and electroweak contributions to the neutral Drell Yan process.

\end{abstract}

\maketitle

\section{Introduction}

\label{sec:intro}

The measurement of the neutral Drell-Yan  (DY) process, $pp\rightarrow Z^*,\gamma^*\rightarrow l^+l^-$, has provided important validation of Standard Model (SM) predictions, has served as a testing ground for searches for high mass $Z^\prime$ bosons and other new physics scenarios, and most recently as a probe of deviations from the SM in an 
effective field theory context.  For all of these applications, precise theoretical predictions both in the SM and in the effective field theory  are crucial.

The SM results for the neutral DY process, along with NLO QCD\cite{Altarelli:1979ub,KubarAndre:1978uy} and NLO electroweak\cite{Baur:2001ze,Dittmaier:2001ay} corrections were 
derived  many years ago.  
  QCD results at NNLO\cite{Hamberg:1990np,Anastasiou:2003yy,Anastasiou:2003ds,Melnikov_2006,Catani:2009sm,Li:2012wna,Camarda:2019zyx} are known for both the total cross section and for some differential
distributions.  Further QCD results exist to N$^3$LL+NNLO and to N$^3$LO\cite{Duhr:2020seh,Duhr:2020sdp,Becher:2020ugp,Re:2021con,Bizo_2018}.
The combined NNLO QCD and NLO  electroweak (EW)
corrections to high mass DY pairs have been studied in detail\cite{Kilgore:2011pa,Buccioni:2020cfi,Delto:2019ewv,Boughezal:2013cwa,Buonocore:2019puv,Dittmaier:2020vra,Bonciani:2020tvf,Heller:2020owb}.  
The state of the art DY predictions are in excellent agreement with experimental results\cite{atlascollaboration2020measurement,Sirunyan:2018owv} suggesting that possible new physics affecting DY production
is either at a very high energy scale or is extremely weakly coupled such that current experiments are only weakly sensitive to these effects.  
 In the effective field theory context, the new physics effects can show up as enhancements at large partonic
energy scales, where  the effects of electroweak
Sudakov logarithms are relatively large\cite{Denner_2001,Campbell:2016dks}, mandating precision calculations  in the
effective field theory beyond the existing SM results.

Without the discovery of new high mass particles, the search for beyond the standard model (BSM)  physics can be pursued using an effective field theory.  
The SM effective field theory (SMEFT)\cite{Brivio:2017vri}  assumes that the Higgs particle is contained in an $SU(2)_L$ doublet and that weak scale interactions
can be described by the Lagrangian,
\begin{equation}
L\sim L_{SM}+\Sigma_{i,n}{C_i^n\over \Lambda^{n-4}}O_i^n\,   ,
\label{eq:ldef}
\end{equation}
where the operators $O_i^n$ have dimension-$n$ and contain only SM particles and  $\Lambda$ parameterizes the ultraviolet (UV)  cut-off scale.  BSM physics is then described by non-zero values of the coefficient
functions $C_i^n$.  Since the operators have dimension greater than $4$, they typically generate effects that grow with energy and can be searched for
in the tails of distributions.

There has been considerable progress in the development of simulation tools for the SMEFT\cite{Brivio:2020onw,Degrande:2020evl}.  At present, the SMEFT dimension-6 operators can be included at NLO QCD using these tools.  There are also numerous
specialized studies of individual processes that include NLO QCD\cite{Baglio:2020oqu,Baglio:2019uty,Baglio:2018bkm,Alioli:2018ljm}.  The  NLO electroweak (EW)  corrections, however, are currently performed on a case by case basis.  Most of the NLO EW studies involve decays:
$H\rightarrow b {\overline b}$\cite{Cullen:2020zof,Cullen:2019nnr,Gauld:2016kuu}, $H\rightarrow \gamma\gamma$\cite{Hartmann_2015,Hartmann_2015x,Dawson:2018liq,Dedes:2018seb}, $H\rightarrow ZZ^*$\cite{Dawson:2018pyl},  $H\rightarrow Z\gamma$\cite{Dawson:2018pyl,Dedes:2019bew}, $H\rightarrow WW^*$\cite{Dawson:2018liq}, $Z\rightarrow f {\overline f}$\cite{Dawson:2019clf,Hartmann:2016pil}, and $t\rightarrow Wb$\cite{Boughezal:2019xpp}.  The only $2\rightarrow 2$ particle scattering process that has been
studied at the NLO EW level is DY.  Our previous DY study\cite{Dawson:2018dxp} concentrated on the effect of a single operator, while the
current study represents the first NLO EW study of a $2\rightarrow 2$ process that includes the effects of 
multiple operators.

At tree level in the dimension-6 truncation of the SMEFT, the DY process  depends 
on operators that affect the input parameter relationships and 
on 4-fermion operators that can dominate the rate at high energy and distort the shapes of kinematic 
distributions\cite{Panico:2021vav,Torre:2020aiz,deBlas:2013qqa}.   
The $Z$ pole resonance contributions  to lepton pair production at NLO QCD and NLO EW  in the SMEFT similarly  involve many additional coefficients
beyond those occurring at tree level\cite{Dawson:2019clf}.  In this work, we extend our previous  DY calculation\cite{Dawson:2018dxp} to include the complete set of SMEFT bosonic operators\cite{Wells:2015uba} that contribute at 
NLO QCD and EW  to the 
process, $q {\overline{q}}\rightarrow \gamma^*,Z^*\rightarrow l^+l  ^-$.   
Our results are of  particular interest in  the low energy limit of UV models that do not 
generate $4$-fermion operators at tree level.  This interesting class of models includes models with
$SU(2)$ scalar singlets, doublets and triplets, as well as models with vector-like fermions\cite{deBlas:2017xtg}. If $4-$fermion operators arise at tree level  (as is the case in models with a heavy $Z^\prime$ boson), the tree level SMEFT effects  from these operators will likely dominate over the NLO EW loop effects.  We are therefore motivated by the case where
$4-$ fermion operators are not generated at  the UV scale and  where the NLO EW effects may play a significant role in low energy  DY phenomenology.   We emphasize, however, that our results are independent of model assumptions and represent an important step in the NLO EW SMEFT program.

In Sect. \ref{sec:basics}, we review the SMEFT formalism relevant for this study, along with the lowest order SMEFT
result for the DY process.  Sect. \ref{sec:nlo}  contains the details of our SMEFT calculation and 
our results are summarized in Sect. \ref{sec:res}. Our complete analytic result is attached as supplemental material which can be included in
 existing Monte Carlo programs. 

\section{SMEFT basics} 
\label{sec:basics}

The SMEFT Lagrangian contains an infinite tower of
 $SU(3)\times SU(2)_L\times U(1)_Y$ invariant  operators
constructed from SM fields.  In this work, we restrict ourselves to the dimension -6 operators, 
assume all coefficients are real and do not consider the effects of CP violation.
We use the Warsaw basis \cite{Buchmuller:1985jz,Grzadkowski:2010es} and normalize the
coefficients as in Eq. \ref{eq:ldef}.
  We include flavor indices on our results  and compute amplitudes
  to linear order in the dimension-6 SMEFT coefficients and at 1-loop in the QCD and electroweak couplings.  
The electroweak sector is described by three input parameters which we take to be $M_W, M_Z$ and $G_\mu$, while the  
electromagnetic coupling, $\alpha$, is  a derived quantity.   We define  $w=M_W^2$, $z=M_Z^2$ and
$v$ to be the $(VEV)^2$ of the Higgs field.\footnote{{\bf{Note our unconventional definition of $v$!}}}

The input parameters are related to the  gauge couplings ${\overline g}_1$
and ${\overline {g}}_2$  in the Lagrangian  to ${\cal {O}}\biggl({1\over \Lambda^2}\biggr)$ as,
 \cite{Dedes:2017zog,Alonso:2013hga,Brivio:2017bnu}
\begin{eqnarray}
w&=&{{\overline g_2^2}v\over 4}  \nonumber \\
z&=&\frac{({\overline g}_1^2+{\overline g}_2^2) v}4+\frac{v^2}{\Lambda^2}\left(\frac18 ({\overline g}_1^2+{\overline g}_2^2) C_{\phi D}+\frac12 {\overline g}_1{\overline g}_2C_{\phi WB} \right).
\end{eqnarray}
Dimension-6 4-fermion operators    give contributions to the decay of the $\mu$, changing the relation between the 
vev-squared, $v$, and $G_\mu$,
\begin{eqnarray}
G_\mu
\equiv \frac1{\sqrt{2} v}-\frac1{2\sqrt{2}\Lambda^2}\biggl( C_{ll,2112}+C_{ll,1221}\biggr)
+{\sqrt{2}\over 2\Lambda^2}\biggl(C_{\phi l,11}^{(3)}+C_{\phi l,22}^{(3)}\biggr)\, ,
\label{eq:gdef}
\end{eqnarray}
where the subscripts refer to the generation.   When we perform our NLO calculations, $\sqrt{v}$ is always defined
as the minimum of the potential, as in Ref. \cite{Degrassi:2014sxa}.

In the dimension-$6$ truncation of the SMEFT, the amplitude  for the DY process can be written  to $1-$loop order as 
\begin{eqnarray}
A & \sim & A_{SM}+\Sigma_i{C_i^6\over \Lambda^2}A_{i,LO}^6
+\Sigma_j {D_j^6\over 16\pi^2\Lambda^2} A_{j,NLO}^6\, ,
\end{eqnarray}
where $A_{SM}, A_{i,LO}^6$, and $A_{j,NLO}^6$ are the SM, dimension-$6$ tree level, and dimension-$6$
$1-$ loop contributions, respectively. We define the linear SMEFT result as,
\begin{eqnarray}
\mid A\mid^2_{lin}&\equiv  & \mid A_{SM}\mid^2
+2 Re\biggl(\Sigma_i A_{SM}^* {C_i^6\over \Lambda^2}A_{i,LO}^6\biggr)
\nonumber \\ && +2Re\biggl(
\Sigma_i A_{SM}^* {D_i^6\over 16 \pi^2  \Lambda^2}A_{j,NLO}^6\biggr) \, . 
\label{eq:linres}
\end{eqnarray}
We note that Eq. \ref{eq:linres} is not positive definite.  For our purposes, we define the quadratic
SMEFT result as 
\begin{eqnarray}
\mid A\mid ^2_{quad}  & \equiv & \mid A_{SM}+\Sigma_i{C_i^6\over \Lambda^2}A_{i,LO}^6
+\Sigma_j {D_j^6\over 16\pi^2\Lambda^2} A_{j,NLO}^6 \mid ^2 \, .
\label{eq:quadres}
\end{eqnarray}
The quantity defined in Eq. \ref{eq:quadres} is what is typically used in SMEFT
 phenomenology studies and in global fits.  There are, however, two types of ``quadratic" ${\cal{O}}({1\over \Lambda^4})$ terms that are not included 
in Eq.  \ref{eq:quadres}.  These are the interference of the dimension-$8$ operators with the SM result\cite{Boughezal:2021tih,Boughezal:2021tih} and the double insertions of the dimension-$6$ operators in the amplitude that are beyond the scope of current NLO EW SMEFT calculations.

\subsection{LO Drell Yan Results}
We write  the helicity amplitudes for $q(p_1)  {\overline q}(p_2) \rightarrow l^+(p_3)l ^-(p_4)$ 
in terms of  the matrix elements,
\begin{equation}
M_{XY}=\biggl[{\overline u}(p_2)\gamma_\mu P_Xu(p_1)\biggr]\cdot \biggl[{\overline u}(p_3)\gamma ^\mu P_Y u(p_4)\
\biggr]\, ,
\end{equation}
with $P_{L,R}={1\mp \gamma_5\over 2}$.  The tree level helicity amplitudes are,
\begin{equation}
A_{XY,LO}=G_{XY}M_{XY}
\end{equation}
with 
\begin{equation}
G_{XY}=G_{XY}^{SM}+\delta G_{XY}. 
\label{eq:treedy}
\end{equation}
Summing over helicity amplitudes and averaging over the spin and color,
\begin{eqnarray}
\mid {\overline A}_{LO}(s,t)\mid^2&\equiv & {1\over 12} \Sigma_{XY}\mid G_{XY}\mid^2\mid M_{XY}\mid^2
\nonumber \\
&\equiv & {1\over 12} \mid A_{LO}\mid^2
\, .
\label{eq:lo}
\end{eqnarray}
The spin and color averaged partonic cross sections are,
\begin{eqnarray}
{d{\hat\sigma}\over dt}&=&{1\over 48 \pi s^2}\biggl\{
(\mid G_{LL}\mid^2 +\mid G_{RR}\mid^2)u^2+(\mid G_{LR}\mid^2 +\mid G_{RL}\mid^2)t^2\biggr\}\nonumber \\
{\hat{\sigma}}_{LO}&=& {1\over 16\pi s^2}\int_{-s}^0 dt \mid {\overline{A}}_{LO}(s,t)\mid^2\, ,
\label{eq:losig}
\end{eqnarray}
where $s=(p_1+p_2)^2, t=(p_1-p_3)^2$ and $u=(p_1-p_4)^2$. 

The SM contribution is, 
\begin{eqnarray} 
G_{LL}^{SM}&=&{4\sqrt{2}w (z-w)\over z}{G_\mu Q_q Q_l \over s}+{g_L^q g_L^l\over s-M_Z^2}\nonumber \\ 
G_{LR}^{SM}&=&{4\sqrt{2}w (z-w)\over z}{G_\mu Q_q Q_l\over s}+{g_L^q g_R^l\over s-M_Z^2}\nonumber \\ 
G_{RL}^{SM}&=&{4\sqrt{2}w (z-w)\over z}{G_\mu Q_q Q_l\over s}+{g_R^q g_L^l\over s-M_Z^2}\nonumber \\ 
G_{RR}^{SM}&=&{4\sqrt{2}w (z-w)\over z}{G_\mu Q_q Q_l\over s}+{g_R^q g_R^l\over s-M_Z^2}\, . 
\label{eq:coupdef}
\end{eqnarray}
with $g_L^f=\sqrt{\sqrt{2}G_\mu z}\biggl[\tau^f-2Q_f\biggl(1-{w\over z}\biggr)\biggr]$, $\tau^f=\pm 1$,  $g_R^f=\sqrt{\sqrt{2}G_\mu z}\biggl[-2Q_f\biggl(1-{w\over z}\biggr)\biggr]$, and $Q_f$ is the fermion charge.

The tree level SMEFT contribution is given in Appendix \ref{sec:lo}
and depends on the coefficients,
\begin{eqnarray}
&& \TphiWB,\, \TphiD,\, \C_{\phi l,11}^{(3)},\,   \C_{\phi l, 22}^{(3)},\,  \C_{\phi l, 22}^{(1)}, \C_{\phi e, 22}, 
 \C_{\phi q, 11}^{(3)}, \,\C_{\phi q, 11}^{(1)}, \,  \C_{\phi u, 11} , \, \C_{\phi d, 11}, \,
 \nonumber \\
 && \biggl[ \, \C_{ll,1221}, \, \C_{ll,2112}, \,
\C_{lq, 2211}^{(3)}, \,  \C_{lq,2211}^{(1)}, \,  
 \C_{qe,1122},\, 
  \C_{lu,2211},\,\C_{ld,2211}, \,\C_{ed, 2211}, \,\C_{eu, 2211},
  \biggr] 
\label{eq:lo_ops}
\end{eqnarray}

The subscripts are generation indices, and the coefficients in the square brackets are 4-fermion operators. 
The numerical impact of the  tree level $4-$fermion operators has been explored in many places\cite{deBlas:2013qqa,Berthier:2015gja,Carpentier:2010ue,Falkowski:2017pss,Cirigliano:2012ab}.
In this work, we will ignore the $4-$fermion contributions and focus instead in the impact of the sub-leading (in $s$)
universal coefficients that involve the bosonic operators at $1-$loop.  Our goal is to begin the exploration of the
effects in Drell Yan production of operators that first arise at $1$-loop order. Typically, the $4-$fermion and bosonic operators 
occur in different types of UV theories.
A complete classification of the quantum numbers of high scale particles that generate the various operators at tree level is given in 
Ref. \cite{deBlas:2017xtg}.
 For example, models with UV scalars typically do not generate $4$-fermion operators\cite{Dawson:2017vgm,deBlas:2014mba}, while a model with a sequential $Z^\prime$  gauge boson  will generate  such  operators\cite{deBlas:2017xtg} that can be probed using kinematic observables in DY production \cite{ Alioli:2017nzr,Panico:2021vav}.

It has been pointed out in Ref. \cite{Breso-Pla:2021qoe} that due to the large 
Drell Yan cross section, information not from the  high energy tail   can potentially yield important information
on SMEFT cross sections and so the coefficients of the  dimension-$6$ operators could be significantly restricted by Drell
Yan scattering, even at relatively low energies.  In fact, the one-loop EW contributions to $Z$  decays in the SMEFT can be as large as ${\cal{O}}(10-20)\%$ at LHC energies\cite{Dawson:2019clf,Dawson:2018jlg}.  Ref. \cite{Farina:2016rws} computed a set of  SMEFT contributions to the  neutral DY process and found
that the high luminosity  LHC will have significant sensitivity to these effects.  Here we 
extend that calculation to include the full set of SMEFT QCD and electroweak corrections from bosonic operators.

\section{NLO Calculation of Drell-Yan Production in the SMEFT}
\label{sec:nlo}
\subsection{Virtual Contributions to NLO SMEFT Drell Yan Results}

\begin{figure}
  \centering
  \hskip -.5in
\includegraphics[width=0.6\textwidth]{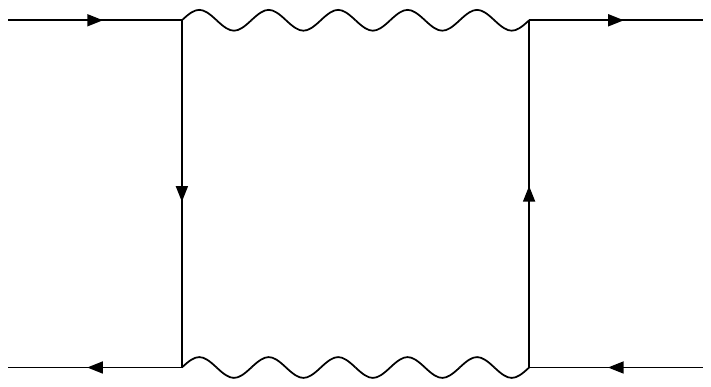}
\hskip -2.5in
\includegraphics[width=0.6\textwidth]{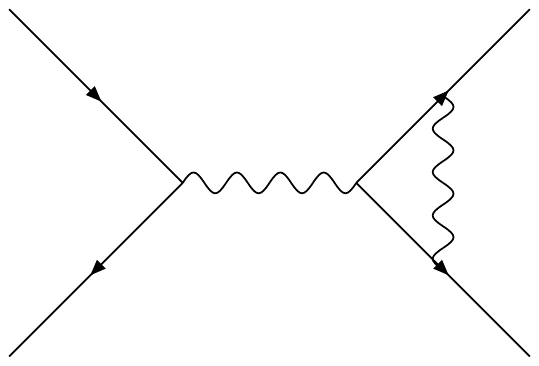}
\hskip -2.3in
\includegraphics[width=0.6\textwidth]{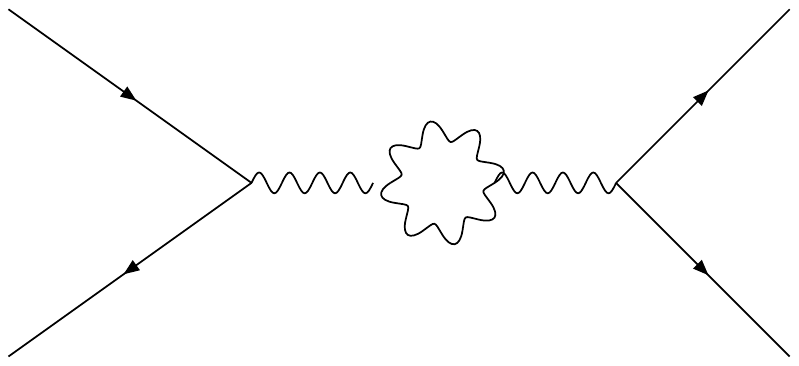}
\vskip -3.5in
 \caption{Sample Feynman diagrams contributing to the $1$-loop virtual contribution.}
   \label{fig:fd}
\end{figure}
In this section we detail the calculation of the virtual NLO corrections to the Drell-Yan process in the SMEFT.  
We follow the notation of\cite{Dawson:2018pyl,Dawson:2019clf,Dawson:2018dxp,Dawson:2018liq}.
The operators that appear in $1-$loop amplitudes are those in Eq. \ref{eq:lo_ops} together with
\begin{eqnarray}
&& \TW,\, \Tphik,\,\TphiW,\,\TphiB, \,\C_{uW,33}, \,\C_{uB,33},\, \C_{\phi l,aa}^{(3)},\,  \C_{\phi l, aa}^{(1)}, \C_{\phi e, aa}, 
 \C_{\phi q, aa}^{(3)}, \,\C_{\phi q, aa}^{(1)}, \,  \C_{\phi u, aa} , \, \C_{\phi d, aa},
 \label{eq:nlo_ops}
\end{eqnarray}
where the subscript $aa=11,22,33$ is a generation index.

It is convenient to separate the contributions to the virtual NLO corrections into box contributions, vertex contributions, propagator contributions and the contributions from renormalization counterterms.  These are shown schematically in Fig. \ref{fig:fd}. 
As we clarified in the introduction, we do not include $4$-fermion operators in our calculation, since we are  interested in effects which first occur at $1-$loop. Furthermore, we calculate only the NLO corrections that interfere with the LO amplitudes.   As a result of these restrictions, the only topologies that enter in our calculations are the ones already present at the SM level. 

The virtual corrections to Drell Yan suffer from UV divergences, along with soft and collinear infrared  (IR) divergences. 
We regularize the UV divergences by working in $d=4-2\epsilon$ dimensions, while the IR divergences are regularized with the introduction of small fermion masses  ($m_f$) and infinitesimal masses for the photon ($m_\gamma$)  and the gluon ($m_g$)~\cite{Baur:1998kt,Dittmaier:2009cr}.
This approach has the advantage of clearly separating the divergences according to their origins and proves to be particularly advantageous in the calculation of the box contributions. 

We extract the box contributions by contracting the NLO one-particle-irreducible (1PI) amplitudes with $M_{XY}$ since in the limit of massless fermions, these matrix elements act as projectors. However, the presence of $\gamma_5$ requires  particular care when working in $d\neq 4$ dimensions. As is well known, $\gamma_5$ is a fundamentally $4-$dimensional object and is not well defined in $d\neq 4$ dimensions, where it is  generally necessary to introduce a  scheme to perform traces involving $\gamma_5$ matrices in a consistent way (e.g. \cite{tHooft:1972tcz,Breitenlohner:1977hr,Larin:1993tq}). The drawback of
these schemes  is the violation of Ward identities, that require the introduction of further counterterms. However, when the traces involve less then $4$ Dirac matrices and a $\gamma_5$,  the results obtained using the Naive Dimensional Regularization (NDR), where the $\gamma_5$ is treated as an anti-commuting object in $d$ dimensions and the Ward identities are conserved, are identical to those obtained using more sophisticated schemes. 

In the case of the SMEFT, the contractions of the 1PI amplitudes with the $M_{XY}$ generate traces involving at least 4 Dirac matrices and a $\gamma_5$, and the NDR cannot be consistently used. However, when the $4$-fermion operators are neglected at $1-$loop and the IR divergences are regularized with finite masses, the box contributions are finite in the limit $d\to 4$. We use these properties to calculate the contractions with the $M_{XY}$ directly in $d=4$, thus avoiding the problem of defining a scheme for the $\gamma_5$. 

The vertex contributions include the 1PI vertex amplitudes and the terms obtained from the fermion wave function renormalization (fWFR).
We extract the 1PI vertex contribution by contracting the NLO amplitudes with the left- and right-handed currents,
 $\biggl[{\overline u}\gamma_\mu P_Lu\biggr]$ and $\biggl[{\overline u}\gamma_\mu P_Ru\biggr]$. Contrary to the box contributions, the vertex contributions are not finite in the limit $d\to 4$. However, since in this case at most three Dirac gammas appear together with a $\gamma_5$, we can rely on the NDR scheme when calculating the vertex contributions.
 It is worth  pointing out, however, that  even though we are ultimately working in the limit of massless fermions, the calculation of the fWFR has to be carried out for massive fermions.
Finally, the propagator contributions can be obtained directly from the gauge boson $2$-point functions in the 
SMEFT \cite{Chen:2013kfa,Ghezzi:2015vva}.

In the renormalization of the LO amplitude, we employ a mixed OS/$\overline{{\rm MS}}$ scheme, where the SM parameters are renormalized in the OS scheme, while the coefficients of the EFT operators are treated as $\overline{{\rm MS}}$ objects.
We use the $\{G_\mu, M_Z, M_W\}$ scheme for the input parameters. The corrections to the masses of the gauge bosons are defined 
 according to 
\begin{equation}
M_V^2=M^2_{0,V}-\Pi_{VV}(M^2_{V}),
\end{equation}
where $V=Z,W$, the $0$ indicates the bare quantities and  $\Pi_{VV}(M^2_{V})$ are the 2-point functions of Refs.  \cite{Chen:2013kfa,Ghezzi:2015vva} computed on-shell.

The relation of Eq.~\ref{eq:gdef} is modified at $1$-loop,
\begin{eqnarray}
G_\mu+ \frac1{2\sqrt{2}\Lambda^2}\biggl( C_{ll,2112}+C_{ll,1221}\biggr)
-{\sqrt{2}\over 2\Lambda^2}\biggl(C_{\phi l,11}^{(3)}+C_{\phi l,22}^{(3)}\biggr)
&\equiv &\frac1{\sqrt{2} v_0}(1+\Delta r)
\nonumber \\
\end{eqnarray}
where  $v_0$ is the {\bf{square}} of the minimum of the potential at tree level and
the analytic expression for $\Delta r$ in the SMEFT is given in Ref.  \cite{Dawson:2018pyl}.

The  effective field theory coefficients of the dimension-6 operators are treated as $\overline{{MS}}$ quantities, defined at the scale of the measurement, i.e. the EW scale.
The poles of the  one-loop coefficients $\C_i$ are extracted from Refs.  \cite{Jenkins:2013zja,Jenkins:2013wua,Alonso:2013hga},
\begin{equation}
\C_i(\mu_R)=\C_{0,i}-\frac1{2\hat{\epsilon}}\frac1{16\pi^2}\gamma_{ij}\C_j,
\end{equation}
where $\mu$ is the renormalization scale,  $\gamma_{ij}$ are the one-loop anomalous dimensions, 
\begin{equation}
\mu_R \frac{d \C_i}{d\mu_R}=\frac1{16\pi^2}\gamma_{ij}\C_j,
\end{equation}
and $\hat{\epsilon}^{-1}\equiv\epsilon^{-1}-\gamma_E+\log(4\pi)$.

It is worth mentioning that, to obtain consistent results in the SMEFT, the definition of renormalizability is modified by requiring that any loop diagram with powers of $1/\Lambda$ higher than the tree level diagrams is set to zero in the calculation. This is done to avoid the appearance of divergences  for which no counterterm can be written without introducing higher order operators at leading order, thus making our renormalization program fail. In our case that means dropping any loop diagram with more than one insertion of dimension 6 operators.

We use the FeynRules routines \cite{Alloul:2013bka} to convert the $R_\xi$ Feynman rules for the SMEFT in the Warsaw basis presented in \cite{Dedes:2017zog} to a FeynArts \cite{Hahn:2000kx} model file. We then compute the amplitudes and reduce the $1$--loop integrals to the Passarino-Veltman \cite{Passarino:1978jh} integrals using FeynCalc \cite{Shtabovenko:2020gxv}.  
The presence of the complicated momentum structure of the SMEFT makes the calculation non-trivial.

\subsection{ Real Contributions to NLO SMEFT Drell Yan Results} 
\label{sec:real}

\label{sec:ap2}
The NLO result requires the real contributions from both photon and gluon emission,
\begin{equation}
q(p_1) {\overline q}(p_2)\rightarrow l^+(p_3) l^- (p_4) \gamma(p_5) ,~q {\overline q}\rightarrow l^+l^- g\, ,
\end{equation}
which gives the spin and color averaged results,
\begin{eqnarray}
&&\mid {\overline A}({\overline q}q\rightarrow l^+\l^-\gamma)\mid^2={1\over 12 }
\cdot\nonumber \\ &&{2\over s_{15}s_{25}s_{35}s_{45}}\biggl\{
s_{12}s_{15}s_{25}Q_l^2~\biggl[ \tilde{t}^2 \biggl(F_{LR}^2(s_{12})+F^2_{RL}(s_{12})\biggr)+ \tilde{u}^2\ \biggl(F^2_{LL}(s_{12}) +F^2_{RR}(s_{12})\biggr)\biggr]\nonumber \\
\nonumber \\ && +s_{34} s_{35} s_{45} Q_q^2 \biggl[\tilde{t}^2 \biggl(F_{LR}^2(s_{34}) +F^2_{RL}(s_{34})\biggr)
 +\tilde{u}^2 \biggl(F^2_{LL}(s_{34})  + F^2_{RR}(s_{34} )\biggr)\biggr]
 \nonumber \\
 && - Q_l Q_q
 \biggl[
 s_{34}^2(t - u) + s_{34}(\tilde{t}^2 - \tilde{u}^2) 
 + {1\over 2}(t + u)(\tilde{u}^2 - \tilde{t}^2 + t^2 - u^2)\biggr]
 \nonumber \\ && \cdot \biggl[ { \tilde{t}^2 }\biggl(F_{LR}(s_{12})F_{LR}(s_{34}) + F_{RL}(s_{12})F_{RL}(s_{34})
 \biggr) 
 \nonumber \\ &&+ 
   { \tilde{u}^2  }\biggl(F_{LL}(s_{12})F_{LL}(s_{34}) + F_{RR}(s_{12})F_{RR}(s_{34})\biggr)
   \biggr]\biggr\}\,
   \label{eq:real}
   \end{eqnarray}
   where $p_1,p_2$ are incoming and $p_3,p_4,p_5$ are outgoing and we define 
   $s_{12}=(p_1+p_2)^2$,  $s_{ij}=(p_i-p_j)^2,~  (i=1,2,~ j=3,4,5)$,  and 
   $s_{jk}=(p_j+p_k)^2,~(j,k=3,4,5)$.
   \begin{eqnarray}
   \tilde{u}^2&=&s_{14}^2+s_{23}^2\nonumber \\
   \tilde{t}^2&=& s_{13}^2 +s_{24}^2\nonumber \\
   u&=& s_{14}+s_{23}\nonumber \\
   t&=& s_{13}+s_{24}
   \end{eqnarray}
The functions $F_{XY}(s_{ij})=4 (\sqrt{\sqrt{2} G_\mu w (z-w)/z}) (F_{XY}^{SM}+\delta F_{XY})$ with $F_{XY}^{SM}= G_{XY}^{SM}$  defined in Eq. \ref{eq:coupdef} and $\delta F_{XY}$ defined in Appendix \ref{sec:realsmeft}.
The SMEFT contributions and the complete real gluon emission contribution can  be found  in Appendix \ref{sec:realsmeft}.
\section{Results}
\label{sec:res}

The IR singularities are regulated using phase space slicing with small photon and gluon masses, $m_\gamma$ and $m_g$, and
also a small fermion mass, $m_f$, as in Refs. \cite{Baur:1998kt,Baur:2001ze,Dittmaier:2001ay,Dittmaier:2009cr}.  After regulating the IR singularities by including the collinear
and soft limits of the $2\rightarrow 3$ real gluon and real photon emission contributions, these masses can be set to $0$.  

The soft limits of the $2\rightarrow 3$
scattering processes have a universal form  that is the same for both the SM and the SMEFT\cite{Denner:2019vbn,denner2007techniques}.  We define the soft
contribution to have  the photon or gluon energy satisfying $E_\gamma, E_{g} < \Delta E$, where $\Delta E$ is an
arbitrary small cut-off.  The soft partonic cross section is defined in terms of the lowest order SMEFT cross section of Eq. \ref{eq:lo},
\begin{eqnarray}
d{\hat\sigma}_{soft}=
&=&{1\over 16\pi s^2}\int_{-s}^0dt \mid {\overline {A}}_{LO}(s,t)\mid^2\biggl(\alpha \delta^{EW} _{soft}(s,t)+\alpha_S C_F
\delta^{QCD}_{soft}(s)\biggr)\, .
\end{eqnarray}
We note that  $\alpha$ is defined in the SMEFT and at this order, the contributions to $\alpha$ come only from the definition of the $Z$ mass and from the $SU(2)/U(1)$ mixing. Notice that there is no modification due, for example, to $C_{\phi W}$. 
\begin{equation}
\alpha=\frac{\sqrt{2}G_\mu w(z-w)}{\pi z}-\frac1{\Lambda^2}\frac{w}{2z\pi}\biggl(w \mathcal{C}_{\phi D} + 4 \sqrt{w(z-w)} \mathcal{C}_{\phi WB}+(z-w)
\biggl[ 2(C_{\phi l,11}^{(3)}+C_{\phi l,22}^{(3)})-C_{ll,1221}-C_{ll,2112}\biggr]\biggr)\, .
\label{eq:asmeft}
\end{equation}

The soft functions are given by, 
\begin{eqnarray}
\delta^{EW}_{soft}&=&Q_q^2 f_q(s)+ Q_l^2 f_l(s)+2Q_q Q_l h(s,t)\nonumber \\
\delta^{QCD}_{soft}&=&f_q(s)\nonumber \\ 
f_f(s)&=& 
 -{1\over \pi}\biggl\{ \ln\biggl({4\Delta E^2\over m_\gamma^2}\biggr)
+\ln\biggl({4\Delta E^2\over m_\gamma^2}\biggr)\log\biggl({m_f^2\over s}\biggr)+\log\biggl({m_f^2\over s}\biggr)
+{1\over 2}\log^2\biggl({m_f^2\over s}\biggr)+{\pi^2\over 3}\biggr\}
\nonumber \\ 
h(s,t)&=& -{1\over \pi}\biggl\{ 
\log\biggl({u\over t} \biggr)\log \biggl({s\over m_\gamma^2}\biggr)
 +\log\biggl({u\over t}\biggr) \log \biggl({4\Delta E^2\over s}\biggr)
 -Li_2\biggl({-t\over u}\biggr)+Li_2\biggl({-u\over t}\biggr)
 \biggr\}
\label{eq:soft}
\end{eqnarray}
For soft gluons, take $Q_l=0$ and $\alpha Q_q^2\rightarrow \alpha_s C_F$.
Adding the virtual one loop contributions  and Eq. \ref{eq:soft}, the $\log(m_{\gamma})$ and $\log(m_g)$ dependences
cancel,  leaving just the $\log(m_f)$ singular terms.

Consider photons emitted from the initial quark within an angle $\delta_\theta$. 
 These give a contribution to the partonic cross section\cite{Harris:2001sx,Baur:1998kt}, 
\begin{eqnarray}
{\hat {\sigma}}^q_{coll} &=& {\alpha\over 2 \pi}Q_q^2\int_0^{1-{2\Delta E/\sqrt{s}} }dy {\hat{\sigma}}_{LO}(y{{s}}) \biggl[{1+y^2\over 1-y}\log\bigg({{ s} \delta_\theta\over 2 m_f^2}\biggr)-{2y\over 1-y}\biggr]\, ,
\end{eqnarray}
plus an identical term for the photon emitted  from the initial ${\overline {q}}$. As usual, $\alpha$ (Eq.\ref{eq:asmeft}) and ${\hat \sigma}_{LO}$ (Eq. \ref{eq:losig}) are defined in the SMEFT.

The initial state collinear contributions to the hadronic cross section are
\begin{eqnarray}
\sigma_{coll}^q&=& {\alpha\over \pi}Q_q^2\int dx_1dx_2 \biggl\{q(x_1){\overline {q}}(x_2)
 \int _0^{1- 2 \Delta E/\sqrt{s}} dy
  {\hat {\sigma}}_{LO}(ys)\nonumber \\ &&
\cdot \biggl[{1+y^2\over 1-y}\log\biggl({{ s}\over m_f^2 }{\delta_\theta\over 2 }\biggr)-{2y\over 1-y}\biggr]
+(1 \leftrightarrow 2)\biggr\}\nonumber \\
&=& 
 {\alpha\over \pi}Q_q^2 \int dx_1 dx_2 \biggl\{q(x_1)  {\hat {\sigma}}_{LO}(s)
 \int _0^{1- 2 \Delta E/\sqrt{s}} {dy\over y}  {\overline {q}}\biggl({x_2\over y}\biggr)
 \nonumber \\ &&
\cdot \biggl[{1+y^2\over 1-y}\log\biggl({{ s}\over m_f^2 }{\delta_\theta\over 2y }\biggr)-{2y\over 1-y}\biggr]
+(1 \leftrightarrow 2)\biggr\}\, ,
\end{eqnarray}
where  in the second equality
we have shifted the argument of ${\hat{\sigma}}_{LO}$ so as to have the partonic center of mass energy be ${{s}}=x_1 x_2 S_H$,
where $S_H$ is the hadronic center of mass energy. 

The initial state collinear contributions are absorbed into the definition of the PDFs\cite{Baur:1998kt}.
In the  ${\overline{MS}}$ scheme, we have
\begin{eqnarray}
q(x)&=& q(x,\mu_F^2)\biggl\{1-{\alpha\over\pi}Q_q^2 \biggr[1
-\log\biggl({2\Delta E\over\sqrt{s}}\biggr)-\log^2\biggl( {2\Delta E\over\sqrt{s}}\biggr)+
\biggl(\log \biggl({2\Delta E\over\sqrt{s}}\biggr)+{3\over 4}\biggr)\ln\biggl({\mu_F^2\over m_f^2}\biggr)\biggr]\biggr\}
\nonumber \\  &&
-{\alpha\over 2 \pi}Q_q^2\int_x^{1-{2\Delta E\over\sqrt{s}}}{dy\over y} q\biggl({x\over y}\biggr) 
\biggl\{ {1+y^2\over 1-y}\log\biggl({\mu_F^2\over m_f^2(1-y)^2}\biggr)
-{1+y^2\over 1-y}\biggr\}\, ,
\label{eq:pdfdef}
\end{eqnarray}
and $\mu_F$ is the factorization scale.  The QCD contribution is found with the replacement $\alpha Q_q\rightarrow \alpha_s  C_F$.

When the photons are emitted from the final state lepton,
\begin{eqnarray}
{\hat\sigma}^l_{coll} &=& {Q_l^2\over 16\pi s^2}\int_{-s}^0 dt \mid {\overline {A}}(s,t)_{LO}
\mid^2 \alpha \delta c^l(s)\nonumber \\ 
\delta c^l(s)&=& {1\over \pi} \int^{1-{2\Delta E\over\sqrt{s}}}_0dy\biggl[{1+y^2\over 1-y}\log\biggl({s\over m_l^2}{\delta_\theta   y^2 \over 2}\biggr)-{2y\over 1-y}\biggr]\, .
\end{eqnarray}
For small $\Delta E$,
\begin{eqnarray}
\delta c^l &=& {Q_l^2\over \pi}  \biggl\{
{9\over 2} +2\log\biggl({2\Delta E\over \sqrt{s}}\biggr)-{2 \pi^2\over 3}
\nonumber \\
&&-\biggl[{3\over 2}+2\log\biggr({2\Delta E\over \sqrt{s}}\biggr)\biggr]
\log\biggl({s\over m_l^2}{\delta_\theta
\over 2}\biggr)\biggr\}\, .
\end{eqnarray}

We write the final answer as the sum of 4 pieces.  The first starts with the partonic combination 
of variables
\begin{eqnarray} 
\hat{\sigma}_a&=& {1\over 16\pi s^2} {1\over 12} \int_{-s}^0dt \mid {A}_{LO}(s,t) + 
\delta A_{NLO}(s,t)\mid^2\, ,
\end{eqnarray}
where 
\begin{eqnarray} 
\delta A_{NLO}(s,t)&=&\biggl\{ \delta A_{\rm{virt}}(s,t)+\frac12 A_{LO}(s,t)\biggl[
\alpha\delta_{soft}^{EW}(s,t)+\alpha_s C_F\delta_{soft}^{QCD}(s)\nonumber \\
&&+(\alpha Q_q^2 +\alpha_s C_F)\delta_{PDF}(s)+\alpha \delta c^l(s)\biggr]\biggr\}\,
\label{eq:virtdef}
\end{eqnarray}
and  $\delta A_{virt}(s,t)$ is the one loop renormalized amplitude calculated in Sect. \ref{sec:nlo} and as always
$A_{LO}$ and $\alpha$ are SMEFT quantities. We note that Eq. \ref{eq:virtdef} is a finite object and
we are free to apply it at linear or quadratic order in the SMEFT following Eqs. \ref{eq:linres} and \ref{eq:quadres}.

The term  $\delta_{PDF}$  arises from the contribution of Eq. \ref{eq:pdfdef} that is proportional to the LO partonic cross section,
\begin{eqnarray}
\delta_{PDF}&=&\biggl(-{2\over \pi}\biggr)\biggl\{ 1-\log\biggl({2\Delta E\over\sqrt{s}}\biggr)
-\log^2\biggl({2\Delta E\over\sqrt{s}}\biggr)
\nonumber \\ && 
+\biggl(\log\biggl({2\Delta E\over\sqrt{s}}\biggr)+{3\over 4}\biggr)
\log\biggl({\mu_F^2\over m_f^2}\biggr)\biggr\}\, .
\end{eqnarray}
Numerical results for $\delta A_{NLO}$ are given in the auxilliary material as a fortran code using
the QCDLoop\cite{Carrazza:2016gav} notation for the one loop integrals. This can be included in existing Monte Carlo codes and is the major result of this paper.  
Note that $d\hat{\sigma}_a$ has no dependence on $m_\gamma$, $m_{g}$ or $m_f$.
$d\hat{\sigma}_a$ contributes to the hadronic cross section,
\begin{eqnarray}
\sigma_A&=& \int dx_1 dx_2 \biggl[q(x_1,\mu_F){\overline{q}}(x_2,\mu_F)\hat{\sigma}_a+(1\leftrightarrow 2)\biggr]
\, .
\end{eqnarray}

The second class of contribution comes from the mass factorization of the PDFs,
\begin{eqnarray}
\sigma_B &=&\int dx_1dx_2\biggl\{  {\hat \sigma}_{LO}({ s})
\int_0^{1-2\Delta E/\sqrt{s}}{dy\over y} \biggl[q(x_1,\mu_F){\overline q}\biggl({x_2\over y},\mu_F
\biggr)
+{\overline{q}}(x_1,\mu_F){q}\biggl({x_2\over y},\mu_F
\biggr)
\biggr]
\nonumber \\ &&\cdot \biggl(\alpha Q_q^2+\alpha_s C_F\biggr) {1\over \pi}  \biggl[
{1+y^2\over 1-y}\log\biggl({s(1-y)^2\over \mu_F^2 y}{\delta_\theta\over 2} \biggr)
+1-y \biggr]+(1\leftrightarrow 2)\biggr\}\, ,
\end{eqnarray}

The next contribution is the hard, non-collinear contribution from the $2\rightarrow 3$ process,
\begin{eqnarray}
\sigma_C&=&\int dx_1 dx_2 q(x_1,\mu_F){\overline{q}}(x_2,\mu_F)\biggl[\sigma(q {\overline q}\rightarrow
l^+l^-\gamma)[E_\gamma>\Delta E, \theta>\delta_\theta]\nonumber \\ &&
+\sigma(q {\overline q}\rightarrow
l^+l^-g)[E_g>\Delta E, \theta>\delta_\theta\biggr]
\end{eqnarray}
where $\theta $ is the angle between the outgoing photon or gluon and the relevant fermions.  
The combination $\sigma_A+\sigma_B+\sigma_C$ is independent of $\delta_\theta$ and $\Delta E$.
 
Finally, there is the contribution from the $2\rightarrow 3$ processes  $g q ({\overline q}) \rightarrow
l^+l^- g$.  These can be found by crossing from the real amplitudes given in the text and in Appendix \ref{sec:realsmeft}.  We neglect the SMEFT contributions from initial state photons, as these effects are highly suppressed.   As a check of our calculation, we have
reproduced the well known SM electroweak and QCD NLO corrections to DY production.

\begin{figure}
  \centering
\includegraphics[width=0.47\textwidth]{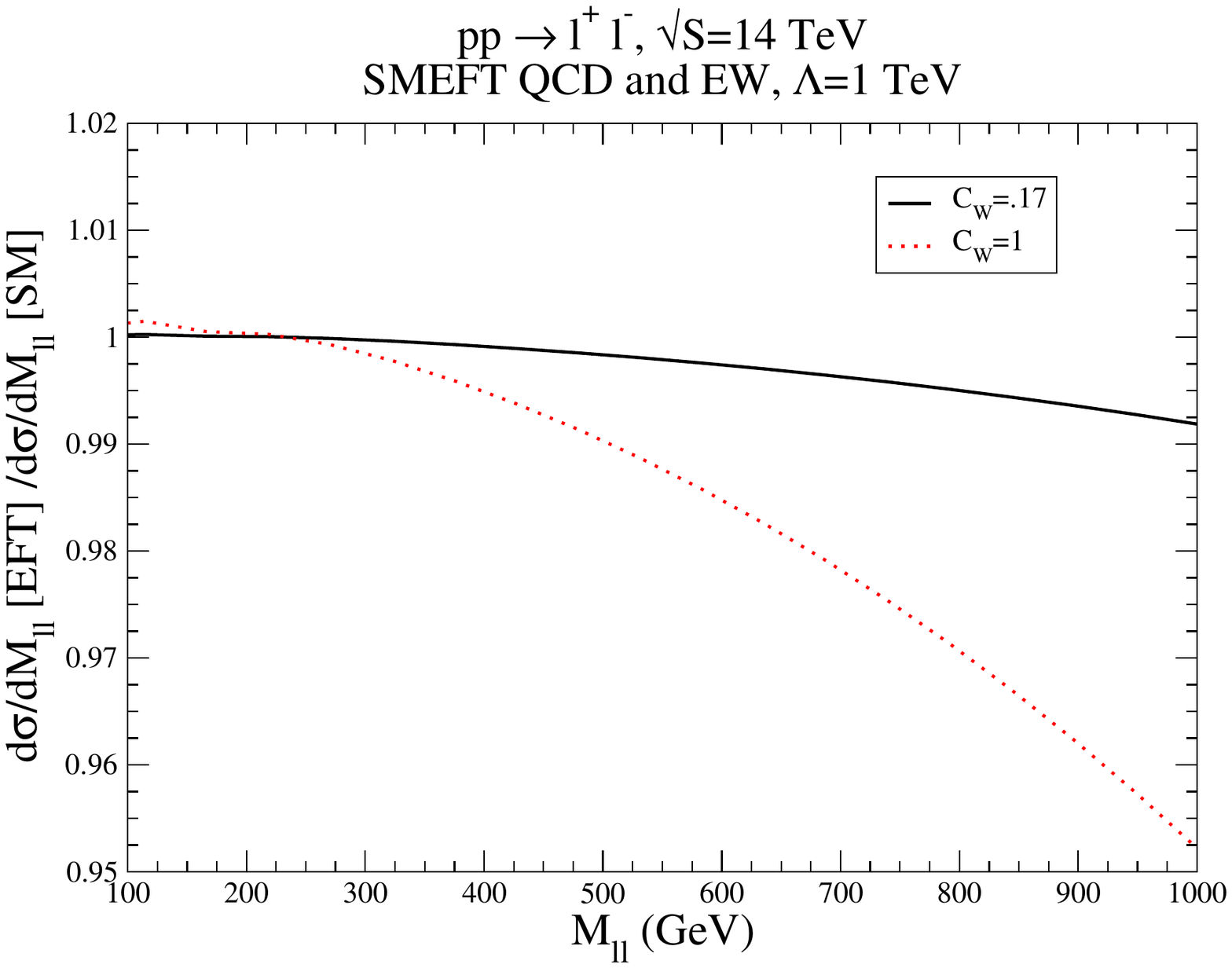}
\includegraphics[width=0.47\textwidth]{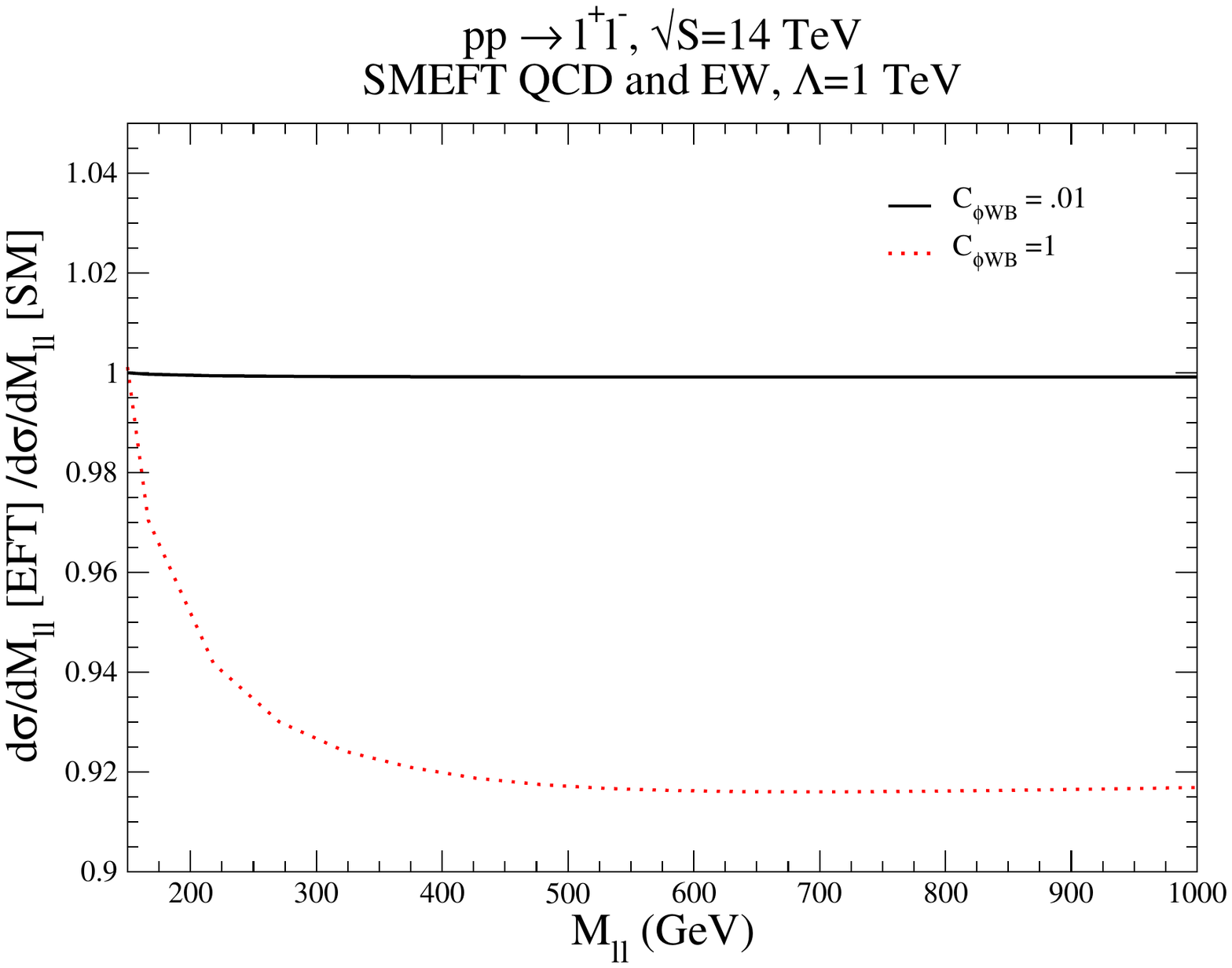}
 \caption{NLO QCD and EW SMEFT contributions to neutral DY production, normalized to the SM NLO predictions.}
   \label{fig:cW}
\end{figure}
\begin{figure}
  \centering
\includegraphics[width=0.4\textwidth]{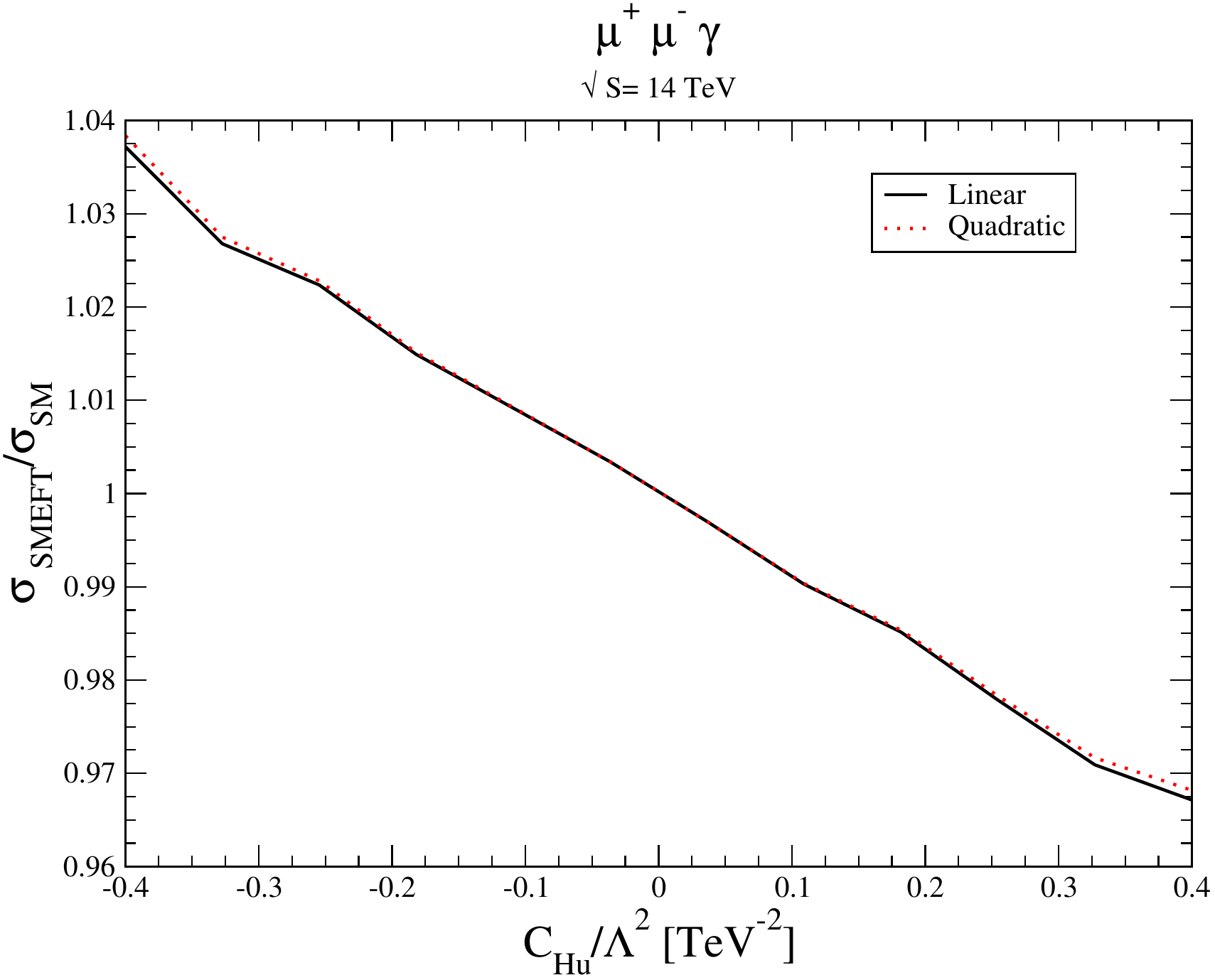}
\hskip .1in
\includegraphics[width=0.38\textwidth]{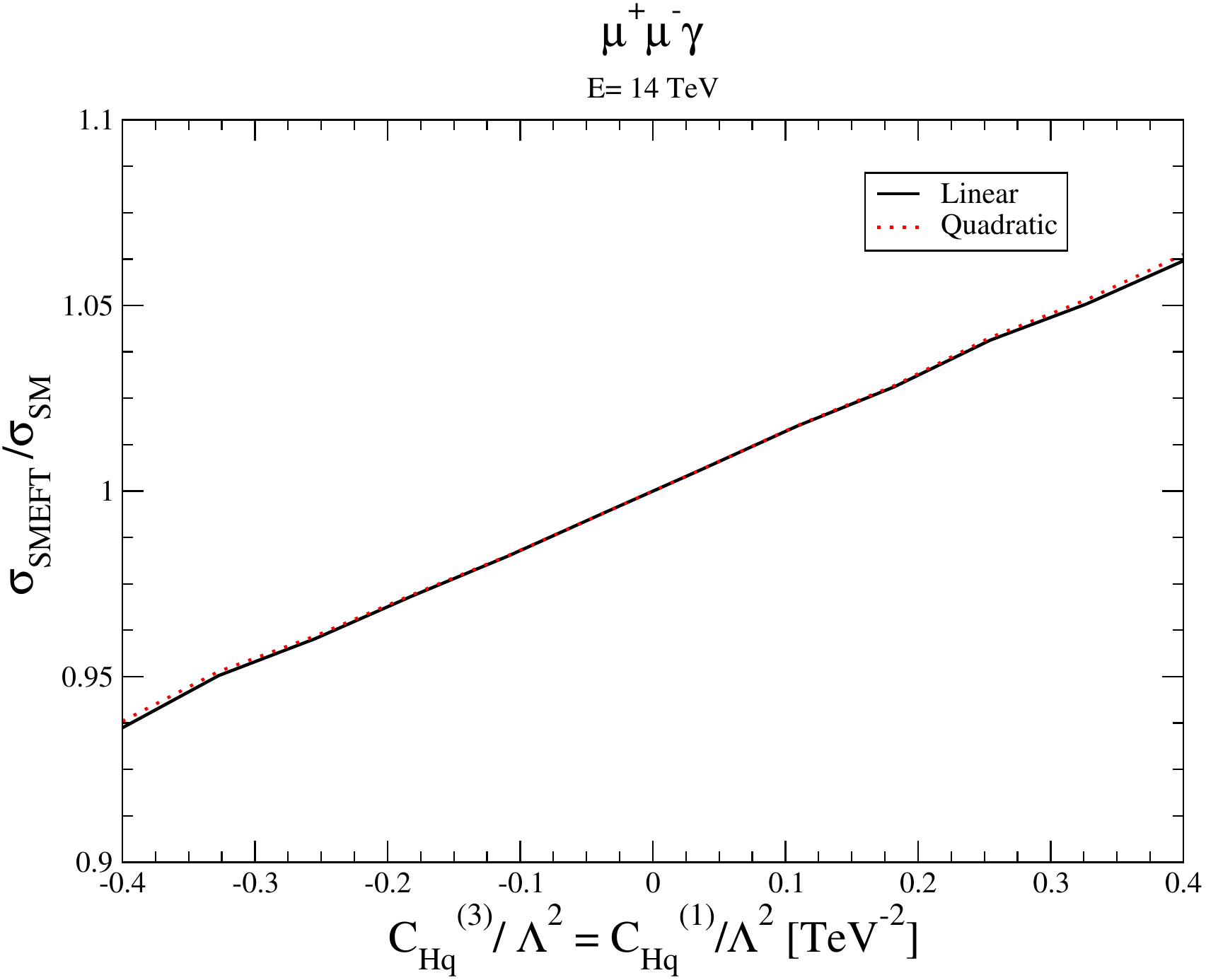}
 \caption{NLO QCD and EW SMEFT contributions to neutral DY production with a tagged $\gamma$, normalized to the SM NLO predictions,
 with $p_T>2 ~GeV$ with the linear and quadratic approximations defined as in Eqs. \ref{eq:linres} and \ref{eq:quadres}. }
   \label{fig:chu}
\end{figure}

In Fig. \ref{fig:cW} we show the effects of $\TphiW$ and $\TphiWB$ at NLO.   Note that the values of $\TphiW$ and $\TphiWB$ that
we have used are larger than those allowed by recent global fits\cite{Alasfar:2020mne,Ethier:2021bye}, emphasizing the smallness of these effects. (An NLO fit to EWPOs\cite{Dawson:2019clf}
gives the allowed $95\%$ CL ranges, $-.0079 < \mathcal{C}_{\phi WB}<.0016$ and $-4.8 < \TW < .48$). The effects of the $C_W$ NLO corrections to DY are  a few percent for allowed values of the coefficients.
As an example, in Fig. \ref{fig:chu} we show the size of the NLO EW effects for the case with tagged photons with $p_T> 2~GeV$ for several coefficients that
are poorly constrained by the global fits.  The size of the effects is of the order of a few percent.  It is interesting that the inclusion of the
quadratic terms has a small effect.  This is consistent with the results of \cite{Ethier:2021bye}.

\label{sec:pheno}

\section{Conclusions}

We have calculated the complete set of NLO electroweak and QCD corrections in the SMEFT to Drell Yan production arising
from bosonic operators, such as those arising in UV models with high scale vector-like fermions or scalars.
The calculation of the virtual EW corrections represents a significant advance in the program of
computing NLO EW effects for scattering processes in the SMEFT.
Our major results are contained in the auxilliary files posted 
at \url{https://quark.phy.bnl.gov/Digital_Data_Archive/dawson/drellyan_21}. The 
results are presented in a form that can be implemented in existing Drell Yan Monte Carlo programs. 
Our results suggest that the NLO corrections from
the bosonic operators are on the order of a few percent in DY processes and presents a target future HL-LHC DY measurements.

\section*{Acknowledgements}

SD is supported by the United States Department of Energy under Grant Contract DE-SC0012704.
The work of PPG has received financial support from Xunta de Galicia (Centro singular de investigaci\'on de Galicia accreditation 2019-2022), by European Union ERDF, and by  ``Mar\'ia  de Maeztu"  Units  of  Excellence program  MDM-2016-0692  and  the Spanish Research State Agency. 

\appendix
\section{Tree level SMEFT results for $q {\overline{q}}\rightarrow l^+ l^-$}
\label{sec:lo}
The tree level SMEFT results  as defined in Eq.  \ref{eq:treedy} for up quark initial states are, 
\begin{eqnarray}
\delta G_{RR}^{u}&=& 
{\TphiWB\over \Lambda^2}  {{16\,\sqrt{w}\,\left(w-s\right)\,\sqrt{z-
 w}}\over{3\,s\,\left(z-s\right)}}
  -{\TphiD\over \Lambda^2}
 {4\,\left(
 s\,z-w^2\right)\over 3\,s\,\left(z-s\right)}
 \nonumber \\ && 
 +{z \over 3\,\left(z-s\right) \Lambda^2 }
 \biggr(
 6\,
 C_{\phi u, 11}-4\,C_{\phi e, 22}+3\,C_{eu, 2211}
 \biggr)
 \nonumber \\ &&
  +{1 \over 3\,\left(z-s\right) \Lambda^2 }
\biggl( \biggl[
 -6\,
 C_{\phi u, 11}+4\,C_{\phi e, 22}\biggr]\,w-3\,C_{eu, 2211}\,s\biggr)
  \nonumber \\
 && +4\frac{(w-s)(z-w)}{3s(z-s)}\biggl(2(C_{\phi l,11}^{(3)}+C_{\phi l,22}^{(3)})-C_{ll,1221}-C_{ll,2112}\biggr)\nonumber \\
\delta G_{LL}^{u}&=&
 {\TphiWB\over \Lambda^2} {{2\,\sqrt{w}\,\left(8\,w-3\,s\right)\,
 \sqrt{z-w}}\over{3\,s\,\left(z-s\right)}}
 -{{\TphiD}\over \Lambda^2}
 {{
 \left(s\,z-8\,w^2\right)}\over   {6\,s\,\left(z-s\right)}}
 \nonumber \\ &&
 -{z
  \over 3\,\left(z-s\right) \Lambda^2} \
 \biggl(
 3\, C_{lq, 2211}^{(3)}-3\,C_{lq,2211}^{(1)}+3\,
 C_{\phi q, 11}^{(3)}-3 C_{\phi q, 11}^{(1)}+C_{\phi l, 22}^{(3)}+
 C_{\phi l, 22}^{(1)}\biggr)
 \nonumber \\ &&
 -{w
  \over 3\,\left(z-s\right)\Lambda^2 } 
 \biggl(
 -6\,C_{\phi q, 11}^{(3)}\,+6\,{C_{\phi q, 11}^{(1)}\,-4
 \,C_{\phi l, 22}^{(3)}-4\,C_{\phi l, 22}^{(1)}\, \biggr)} \nonumber \\
 &&-{s
  \over 3\,\left(z-s\right)\Lambda^2 }
 \biggl(-3\,C_{lq,2211}^{(3)}
 +3\,C_{lq, 2211}^{(1)}\,\biggr)\nonumber \\
 &&-\frac{8w(w-z)+s(2w+z)}{6s(z-s)}\biggl(2(C_{\phi l,11}^{(3)}+C_{\phi l,22}^{(3)})-C_{ll,1221}-C_{ll,2112}\biggr)\nonumber \\
 \delta G_{LR}^{u}&=& 
 {\TphiWB\over \Lambda^2} {{2\,\sqrt{w}\,\left(8\,w-5\,s\right)\,
 \sqrt{z-w}}\over{3\,s\,\left(z-s\right)}}
 -{  \TphiD\over \Lambda^2} \,
 { \left(s\,z-4\,w^2\right) \over {3\,s\,\left(z-s\right)}}
 \nonumber \\ && 
 +{ z \over 3\,\left(z-s\right) \Lambda^2}
 \biggl(
 3\,C_{qe,1122}\,-6 C_{\phi q, 11}^{(3)}\,+6\,
 C_{\phi q, 11}^{(1)}\,-C_{\phi e, 22}\,
 \biggr)\nonumber \\
 && 
 +{ 1 \over 3\,\left(z-s\right) \Lambda^2}\biggl(
 \biggl[6\,C_{\phi q,11}^{(3)}-6\,
 C_{\phi q, 11}^{(1)}+4\,C_{\phi e, 22}\biggr]\,w-3\,C_{qe,1122}\,s
 \biggr)
 \nonumber \\
 &&+\frac{(s-4w)(w-z)}{3s(z-s)}\biggl(2(C_{\phi l,11}^{(3)}+C_{\phi l,22}^{(3)})-C_{ll,1221}-C_{ll,2112}\biggr)\nonumber \\
 \delta G_{RL}^{u}
  &=& 
  {\TphiWB\over\Lambda^2} {4\,\sqrt{w}\,\left(4\,w-3\,s\right)\,
 \sqrt{z-w} \over  3\,s\,\left(z-s\right)}
 -{\TphiD \over \Lambda^2}
 {2\,
 \,\left(s\,z-2\,w^2\right) \over 3\,s\,\left(z-s\right)}
 \nonumber \\ &&
 +{z \over 3\,\left(z-s\right)\Lambda^2}
 \,\biggl(3\,C_{lu,2211}\,+3\,C_{\phi u, 11}\,-4\,
 C_{\phi l, 22}^{(3)}\,-4\,C_{\phi l, 22}^{(1)}\,\biggr)
 \nonumber \\ && 
 +{1 \over 3\,\left(z-s\right)\Lambda^2}
 \biggl(
 \biggl[-6\,C_{\phi u, 11}\,+4\,
 C_{\phi l, 22}^{(3)}\,+4\,C_{\phi l, 22}^{(1)}\,\biggr]w-3\,C_{lu, 2211}\,s\biggr)\nonumber\\
 &&+\frac{2(s-2w)(w-z)}{3s(z-s)}\biggl(2(C_{\phi l,11}^{(3)}+C_{\phi l,22}^{(3)})-C_{ll,1221}-C_{ll,2112}\biggr)
   \end{eqnarray}
   and the numerical subscripts are generation indices. 
 The results for down quark initial states are,   
   \begin{eqnarray}
\delta G_{RR}^{d}&=& 
-{\TphiWB\over \Lambda^2}  {{8\,\sqrt{w}\,\left(w-s\right)\,\sqrt{z-
 w}}\over{3\,s\,\left(z-s\right)}}
  +{\TphiD\over \Lambda^2}
 {2\,\left(
 s\,z-w^2\right)\over 3\,s\,\left(z-s\right)}
 \nonumber \\ && 
 +{z \over 3\,\left(z-s\right) \Lambda^2 }
 \biggr(
 6\,
 C_{\phi d, 11}+2\,C_{\phi e, 22}+3\,C_{ed, 2211}
 \biggr)
 \nonumber \\ &&
  +{1 \over 3\,\left(z-s\right) \Lambda^2 }
\biggl( \biggl[
 -6\,
 C_{\phi d, 11}-2\,C_{\phi e, 22}\biggr]\,w-3\,C_{ed, 2211}\,s\biggr)
  \nonumber \\
 && -2\frac{(w-s)(z-w)}{3s(z-s)}\biggl(2(C_{\phi l,11}^{(3)}+C_{\phi l,22}^{(3)})-C_{ll,1221}-C_{ll,2112}\biggr)\nonumber \\
\delta G_{LL}^{d}&=&
- {\TphiWB\over \Lambda^2} {{2\,\sqrt{w}\,\left(4\,w\right)\,
 \sqrt{z-w}}\over{3\,s\,\left(z-s\right)}}
 -{{\TphiD}\over \Lambda^2}
 {{
 \left(s\,z+4\,w^2\right)}\over   {6\,s\,\left(z-s\right)}}
 \nonumber \\ &&
 +{z
  \over 3\,\left(z-s\right) \Lambda^2} \
 \biggl(
 3\, C_{lq, 2211}^{(3)}+3\,C_{lq,2211}^{(1)}+3\,
 C_{\phi q, 11}^{(3)}+3 C_{\phi q, 11}^{(1)}-C_{\phi l, 22}^{(3)}-
 C_{\phi l, 22}^{(1)}\biggr)
 \nonumber \\ &&
 -{w
  \over 3\,\left(z-s\right)\Lambda^2 } 
 \biggl(
 6\,C_{\phi q, 11}^{(3)}\,+6\,{C_{\phi q, 11}^{(1)}\,+2
 \,C_{\phi l, 22}^{(3)}+2\,C_{\phi l, 22}^{(1)}\, \biggr)} \nonumber \\
 &&+{s
  \over 3\,\left(z-s\right)\Lambda^2 }
 \biggl(-3\,C_{lq,2211}^{(3)}
 -3\,C_{lq, 2211}^{(1)}\,\biggr)\nonumber \\
 &&+\frac{4w(w-z)+s(4w-z)}{6s(z-s)}\biggl(2(C_{\phi l,11}^{(3)}+C_{\phi l,22}^{(3)})-C_{ll,1221}-C_{ll,2112}\biggr)\nonumber \\
 \delta G_{LR}^{d}&=& 
 -{\TphiWB\over \Lambda^2} {{2\,\sqrt{w}\,\left(4\,w-s\right)\,
 \sqrt{z-w}}\over{3\,s\,\left(z-s\right)}}
 -{  \TphiD\over \Lambda^2} \,
 { \left(s\,z+2\,w^2\right) \over {3\,s\,\left(z-s\right)}}
 \nonumber \\ && 
 +{ z \over 3\,\left(z-s\right) \Lambda^2}
 \biggl(
 3\,C_{qe,1122}\,+6 C_{\phi q, 11}^{(3)}\,+6\,
 C_{\phi q, 11}^{(1)}\,-C_{\phi e, 22}\,
 \biggr)\nonumber \\
 && 
 -{ 1 \over 3\,\left(z-s\right) \Lambda^2}\biggl(
 \biggl[6\,C_{\phi q,11}^{(3)}+6\,
 C_{\phi q, 11}^{(1)}+2\,C_{\phi e, 22}\biggr]\,w+3\,C_{qe,1122}\,s
 \biggr)
 \nonumber \\
 &&+\frac{(s+2w)(w-z)}{3s(z-s)}\biggl(2(C_{\phi l,11}^{(3)}+C_{\phi l,22}^{(3)})-C_{ll,1221}-C_{ll,2112}\biggr)\nonumber \\
 \delta G_{RL}^{d}
  &=& 
  -{\TphiWB\over\Lambda^2} {2\,\sqrt{w}\,\left(4\,w-3\,s\right)\,
 \sqrt{z-w} \over  3\,s\,\left(z-s\right)}
 +{\TphiD \over \Lambda^2}
 {
 \,\left(s\,z-2\,w^2\right) \over 3\,s\,\left(z-s\right)}
 \nonumber \\ &&
 +{z \over 3\,\left(z-s\right)\Lambda^2}
 \,\biggl(3\,C_{ld,2211}\,+3\,C_{\phi d, 11}\,+2\,
 C_{\phi l, 22}^{(3)}\,+2\,C_{\phi l, 22}^{(1)}\,\biggr)
 \nonumber \\ && 
 +{1 \over 3\,\left(z-s\right)\Lambda^2}
 \biggl(
 \biggl[-6\,C_{\phi d, 11}\,-2\,
 C_{\phi l, 22}^{(3)}\,-2\,C_{\phi l, 22}^{(1)}\,\biggr]w-3\,C_{ld, 2211}\,s\biggr)\nonumber\\
 &&-\frac{(s-2w)(w-z)}{3s(z-s)}\biggl(2(C_{\phi l,11}^{(3)}+C_{\phi l,22}^{(3)})-C_{ll,1221}-C_{ll,2112}\biggr)
   \end{eqnarray}
   .

\section{Real SMEFT contributions}
\label{sec:realsmeft}
\subsection{Real Photon Emission}
The SMEFT contributions to $q {\overline {q}}\rightarrow l^+ l^-\gamma$  are defined in Eq. \ref{eq:real} and $G_{XY}^{SM}$ is defined in Eq. \ref{eq:coupdef}.
The functions $\delta F_{XY}^q$ are,
\begin{eqnarray} 
\delta F_{XY}^q(s)&=& \frac{f^q_0}{s}  + \frac{f^q_{XY}}{s-z} + \frac{3vG_{XY}^{SM}}{4\Lambda^2}\biggl(C_{ll,1221}+C_{ll,2112}-2(C_{\phi l,11}^{(3)}+C_{\phi l,22}^{(3)})\biggr)
\end{eqnarray}
(Our notation is $w=M_W^2, v=(VEV)^2={1\over\sqrt{2}G_\mu},z=M_Z^2$ as in the main text.)
\begin{eqnarray}
f^u_0& = &{2w\over z\Lambda^2 }\biggl(w \TphiD +4 (z-w) \TphiWBr\biggr)
\nonumber \\
f^u_{LL} &=& {1\over 12z\Lambda^2 }
\biggl\{\TphiD(-24 w^3 + 22 w^2 z + w z^2 - 2 z^3)/(w-z)
%\nonumber \\ &&
  + 4\biggl[\TphiWBr(24w^2 - 28w z + 7z^2) 
\nonumber \\ && +z (z-4w)(
\C_{\phi l 22}^{(1)} 
 + \C_{\phi l 22}^{(3)} ) +3 z (2w-z) (\C_{\phi q 11}^{(1)} - \C_{\phi q 11}^{(3)}) 
 %\nonumber \\ && 
 \biggr]
 \biggr\}
\nonumber \\
f^u_{LR} &=& {1\over 6z\Lambda^2 }
\biggl\{\TphiD(-12 w^2 + w z + 2 z^2)
  + 24\TphiWBr(2w^2 - 3w z + z^2) 
\nonumber \\ && +2 z (z-4w)
\C_{\phi e 22} +12 z (w-z) (\C_{\phi q 11}^{(1)} - \C_{\phi q 11}^{(3)}) 
 \biggr\}
\nonumber \\
f^u_{RL} &=& {1\over 3z\Lambda^2 }
\biggl\{\TphiD(-6 w^2 + w z + 2 z^2)
  + 8\TphiWBr(3w^2 - 5w z +2 z^2) 
\nonumber \\ && +(6 w - 3 z) z
\C_{\phi u 11} +4 z (z-w) (\C_{\phi l 22}^{(1)} + \C_{\phi l 22}^{(3)}) 
 \biggr\}
\nonumber \\
f^u_{RR}& = & {2(z-w )\over 3 z \Lambda^2}
\biggl\{ (3w+2z)\TphiD  + 12(z-w)\TphiWBr w + 2z\Tfebb - 3z\Tfuaa
\biggr\}
\end{eqnarray}

\begin{eqnarray}
f^d_0& = -&{w\over z\Lambda^2 }\biggl(w \TphiD +4 (z-w) \TphiWBr\biggr)
\nonumber \\
f^d_{LL} &=& {1\over 12z\Lambda^2 }
\biggl\{\TphiD(12 w^3 - 8 w^2 z + w z^2 - 2 z^3)/(w-z)
%\nonumber \\ &&
  + 4\biggl[\TphiWBr(-12w^2 +8 w z + z^2) 
\nonumber \\ && +z (z+2w)(
\C_{\phi l 22}^{(1)} 
 + \C_{\phi l 22}^{(3)} ) +3 z (2w-z) (\C_{\phi q 11}^{(1)} + \C_{\phi q 11}^{(3)}) 
 %\nonumber \\ && 
 \biggr]
 \biggr\}
\nonumber \\
f^d_{LR} &=& {1\over 6z\Lambda^2 }
\biggl\{\TphiD(6 w^2 + w z + 2 z^2)
  + 24\TphiWBr w(z-w) 
\nonumber \\ && +2 z (z+2w)
\C_{\phi e 22} +12 z (w-z) (\C_{\phi q 11}^{(1)} + \C_{\phi q 11}^{(3)}) 
 \biggr\}
\nonumber \\
f^d_{RL} &=& {1\over 6z\Lambda^2 }
\biggl\{\TphiD(3 w - 2 z) (2 w + z)
  - 8\TphiWBr(3 w - 2 z) (w - z)
\nonumber \\ && +6(2 w -  z) z
\C_{\phi d 11} +4 z (w-z) (\C_{\phi l 22}^{(1)} + \C_{\phi l 22}^{(3)}) 
 \biggr\}
\nonumber \\
f^d_{RR}& = & {(w-z )\over 3 z \Lambda^2}
\biggl\{ (3w+2z)\TphiD  + 12(z-w)\TphiWBr w + 2z\Tfebb + 6z\Tfdaa
\biggr\}
\end{eqnarray}
where $\TphiWBr\equiv \sqrt{{w\over z-w}}\TphiWB$.

\subsection{Real gluon emission}
The total spin and color averaged amplitude squared for  $q {\overline {q}}\rightarrow l^+ l^-g$  is
\begin{eqnarray} 
\mid {\overline A}\mid^2&=&
{C_F s_{34} \over 6 s_{15}s_{25}} \biggl\{ {\tilde{t}}^2\biggl[ H_{LR}(s_{34})^2 + H_{RL}(s_{34})^2\biggr] 
+ {\tilde{u}}^2\biggl[H_{LL}(s_{34})^2 + H_{RR}(s_{34} ^2)\biggr]
\biggr\}
\end{eqnarray}
with $H_{XY}=2 g_s (H_{XY}^{SM}+\delta H_{XY})$.

The SM contributions are
\begin{equation}
H_{XY}^{SM} (s_{34})=G_{XY}^{SM}(s_{34})
\end{equation}

The SMEFT contributions,  $\delta H_{XY}$, are,
\begin{eqnarray} 
\delta H^q_{XY}(s)&=&\frac{h^q_0}{s}  + \frac{h^q_{XY}}{s-z}+ \frac{vG_{XY}^{SM}}{2\Lambda^2}\biggl(C_{ll,1221}+C_{ll,2112}-2(C_{\phi l,11}^{(3)}+C_{\phi l,22}^{(3)})\biggr)
\end{eqnarray}

\begin{eqnarray}
h^u_0 &=& {4 w\over 3 z \Lambda^2} \biggl\{\TphiD w + 4\TphiWBr(z-w)\biggr\}
\nonumber \\
 h^u_{LL} &=& {1\over 6 z \Lambda^2}\biggl\{
\TphiD (z^2-8w^2) + 4\TphiWBr(8w^2 - 11w z + 3z^2) 
\nonumber \\ &&
 + 2z \biggl[(z-4 w)(\C_{\phi l 22}^{(1)} + \C_{\phi l 2 2}^{(3)}) + 3 (2w-z)(\C_{\phi q 1 1}^{(1)} - \C_{\phi q 1 1 }^{(3)}  ) \biggr] \biggr\}
\nonumber \\ 
 h^u_{LR}& =&{1\over 3 z \Lambda^2} \biggl\{\TphiD(z^2-4w^2) + 2\TphiWBr(8w^2 - 13wz + 5z^2)
\nonumber  \\ &&  + z\biggl[(z-4w) \Tfebb+6(w-z)(\C_{\phi q 1 1}^{(1)} - \C_{\phi q 1 1 }^{(3)} )\biggr]
\nonumber \\
 h^u_{RL}& =& {1\over 3 z \Lambda^2}\biggl\{ \TphiD(2z^2-4w^2) +4\TphiWBr(4w^2 - 7wz + 3z^2) 
\nonumber \\ && + z \biggl[4 (z-w) (\C_{\phi l 2 2}^{(1)} 
 +\C_{\phi l 2 2}^{(3)})  + (6w-3z)\Tfuaa\biggr]\biggr\}
\nonumber\\
 h^u_{RR}& =& {2(z - w)\over  3 z \Lambda^2} \biggl\{
 8\TphiWBr (z-w) + 2 \TphiD (w + z) + 2 z\Tfebb - 3 z \Tfuaa\biggr\}
\biggr\} 
\end{eqnarray}

\begin{eqnarray}
h^d_0 &=& {-2 w\over 3 z \Lambda^2} \biggl\{\TphiD w + 4\TphiWBr(z-w)\biggr\}
\nonumber \\
 h^d_{LL} &=& {1\over 6 z \Lambda^2}\biggl\{
\TphiD (z^2+4w^2) + 16\TphiWBr w (z-w) 
\nonumber \\ &&
 + 2z \biggl[(z+2 w)(\C_{\phi l 22}^{(1)} + \C_{\phi l 2 2}^{(3)}) + 3 (2w-z)(\C_{\phi q 1 1}^{(1)} + \C_{\phi q 1 1 }^{(3)}  ) \biggr] \biggr\}
\nonumber \\ 
 h^d_{LR}& =&{1\over 3 z \Lambda^2} \biggl\{\TphiD(z^2+2w^2) - 2\TphiWBr(4w^2 - 5wz + z^2)
\nonumber  \\ &&  + z\biggl[(z+2w) \Tfebb+6(w-z)(\C_{\phi q 1 1}^{(1)} + \C_{\phi q 1 1 }^{(3)} )\biggr]
\nonumber \\
 h^d_{RL}& =& {1\over 3 z \Lambda^2}\biggl\{ \TphiD(2w^2-z^2) -2\TphiWBr(4w^2 - 7wz + 3z^2) 
\nonumber \\ && + z \biggl[2 (w-z) (\C_{\phi l 2 2}^{(1)} 
 +\C_{\phi l 2 2}^{(3)})  + (6w-3z)\Tfdaa\biggr]\biggr\}
\nonumber\\
 h^d_{RR}& =& {2(w - z)\over  3 z \Lambda^2} \biggl\{
 4\TphiWBr (z-w) + \TphiD (w + z) +  z\Tfebb + 3 z \Tfdaa\biggr\}
\biggr\} 
\end{eqnarray}

%\end{document}
\bibliographystyle{utphys}
\bibliography{DY_paper.bib}

\end{document}